\documentclass[12pt,preprint]{aastex}
\newcommand{\kms}{{\,\rm km\,s}^{-1}}

\def\la{\mathrel{\hbox{\rlap{\hbox{\lower4pt\hbox{$\sim$}}}\hbox{$<$}}}}
\def\sun{\hbox{$\odot$}}
\renewcommand{\mag}{\mbox{$\;$mag}}
\newcommand{\dex}{\mbox{$\;$dex}}

\begin{document}
\title{\uppercase{%
     Temperature Differences in the Cepheid Instability Strip
     Require Differences in the Period-Luminosity Relation in Slope
     and \\ Zero Point}} 

\author{Allan Sandage}
\affil{The Observatories of the Carnegie Institution of Washington,\\
       813 Santa Barbara Street, Pasadena, CA 91101}
\and

\author{G. A. Tammann}
\affil{Department of Physics and Astronomy,\\
       Klingelbergstrasse 82, CH-4056 Basel, Switzerland}
\email{g-a.tammann@unibas.ch}
%
\begin{abstract}
 A graphical and an algebraic demonstration is made to show why the
 slope and zero point of the Cepheid period-luminosity (P-L) relation
 is rigidly coupled with the slope and zero point of the Cepheid
 instability strip in the HR diagram. The graphical demonstration uses
 an arbitrary (toy) ridge line in the instability strip, while the
 algebraic demonstration uses the pulsation equation into which the
 observed P-L relations for the Galaxy and the LMC are put to predict
 the temperature zero points and slopes of the instability strips. 
 Agreement between the predicted and measured  slopes in the
 instability strips argue that the observed P-L differences between
 the Galaxy and LMC are real. In another proof, the direct evidence
 for different P-L slopes in different galaxies is shown by comparing
 the Cepheid data in the Galaxy, the combined data in NGC\,3351 and
 NGC\,4321, in M31, LMC, SMC, IC\,1613, NGC\,3109, and in Sextans A+B. 
 The P-L slopes for the Galaxy, NGC\,3351,  NGC\,4321, and M31 are
 nearly identical and are the steepest in the sample. 
 The P-L slopes decrease monotonically with metallicity in the
 order listed, showing that the P-L relation is not the same in
 different galaxies, complicating their use in calibrating the
 extragalactic distance scale.    
\end{abstract}

\section{INTRODUCTION}
\label{sec:01}
There is evidence that the Cepheid period-luminosity relation is not
universal but differs in slope and zero point from galaxy-to-galaxy at
a level of up to $\sim\!0.3\mag$ as a function of period 
(cf.\
\citealt{Tammann:Reindl:02}; 
\citealt{Tammann:etal:02}; 
\citealt{TSR:03}, hereafter \citeauthor*{TSR:03}; 
\citealt{STR:04}, hereafter \citeauthor*{STR:04}; 
\citealt{ST:06} for a review). 
Drastic as this conclusion is for studies of the extragalactic
distance scale, it has been strengthened in confirming studies by 
\citet{Ngeow:etal:03}, 
\citet{Kanbur:Ngeow:04}, 
\citet{Ngeow:Kanbur:04,Ngeow:Kanbur:05,Ngeow:Kanbur:06},
\citet{Ngeow:etal:05}, 
\citet{Koen:etal:07}, 
and from theoretical models as a function of chemical composition by
many authors, starting perhaps with John
\citet{Cox:59,Cox:80}, and including 
\citet{Christy:etal:66,Christy:etal:72}, 
\citet{Iben:Tuggle:75}, 
\citet{Chiosi:etal:92}, and more recently 
\citet{Bono:etal:00}, 
\citet{Fiorentino:etal:02}, 
\citet{Marconi:etal:05}, 
and undoubtedly others. 

     These studies show that the position of the borders of the the 
$L$, $T_{e}$ instability strip in the HR diagram depends on chemical
composition. If the strip borders vary in position and slope, so must
the slope and zero point of the P-L relation, as worked through the
pulsation equation in the following sections.   

     Despite this evidence, the conclusion that different P-L
relations apply in different galaxies has recently been challenged in
the literature. In these papers it is said that the slopes of the
Cepheid P-L relations in other galaxies satisfy the slope of the P-L
relation in the LMC and therefore that no slope differences with LMC
have been demonstrated conclusively 
(cf. \citealt{Gieren:etal:05a,Gieren:etal:05b,Gieren:etal:06}; 
\citealt{Pietrzynski:etal:06}; 
\citealt{Benedict:etal:07};
\citealt{vanLeeuwen:etal:07} are examples).    

     However, this claim sets aside the parallel evidence that the
slope and zero point of the ridge lines of the Cepheid instability
strips of the Galaxy, LMC and SMC themselves differ in temperature at
a given period (cf.\ Fig.~3 of \citeauthor*{STR:04}), and hence, in
luminosity.  

     The purpose of this paper is to again remind us that the slope of
the P-L relation is rigidly coupled with the slope of the instability
strip via the \citet{Ritter:79} pulsation condition that 
$P\sqrt{\rho}=\;$constant. Hence, if the instability strip slope
varies from galaxy-to-galaxy, so must the P-L slope. 

     Differences in the instability strips of the Galaxy and SMC were
first set out by \citet{Gascoigne:Kron:65}. 
They were made secure as temperature differences by 
\citet{Laney:Stobie:86}, and have now been made definitive by the new
CCD data by  \citet{Udalski:etal:99a,Udalski:etal:99b}
for LMC and SMC and by \citet{Berdnikov:etal:00} for the Galaxy, as
summarized for the Galaxy and LMC in Figure~20 of \citeauthor*{STR:04}.         

     In the next section we show the pulsation equation graphically
and demonstrate from it the stated premise; a slope difference in the
ridge line of the instability strip leads to a slope difference in the
P-L relation. The graphical solution here is parallel to the algebraic
demonstration given elsewhere 
(\citeauthor*{TSR:03}, \S~7.3; \citeauthor*{STR:04}, \S~8), 
and made more explicit here in \S~\ref{sec:03}.

\section{A GRAPHICAL SOLUTION BASED ON THE LINES OF CONSTANT PERIOD 
         IN THE HR DIAGRAM}
\label{sec:02}
In an obvious way the Ritter $P\sqrt{\rho}$ pulsation condition 
can be put into the observable parameters of period, luminosity, 
mass, and temperature by also using the Stefan-Boltzmann black 
body radiation condition that $L\sim R^{2} T_{e}^{4}$. The Ritter plus
black body condition is improved by model calculations for real stars 
by using details of the pulsating stellar atmosphere structure, 
leading to the more precise pulsation equation of $P(L,M,T_{e})$. 

     As in previous papers we again use the van Albada-Baker 
\citeyearpar{vanAlbada:Baker:73} pulsation equation. 
Although it was calculated by them to apply to the lower mass 
RR Lyrae stars, comparisons show that their predicted P-L
relation is nearly identical with many other pulsation equations
calculated for higher mass Cepheids. Examples are the equations by 
\citet[][their eq.~$\lbrack$3$\rbrack$]{Iben:Tuggle:75}, 
\citet[][their eq.~$\lbrack$5$\rbrack$]{Chiosi:etal:92}, 
\citet[][their eq.~$\lbrack$2$\rbrack$]{Simon:Clement:93}, and 
\citet{Saio:Gautschy:98}. 
The near identity among the equations is discussed in 
\citet[][hereafter \citeauthor*{SBT:99}]{SBT:99}. 

     The pulsation equation by \citeauthor{vanAlbada:Baker:73} is 
\begin{equation}
     \log P = 0.84 \log L_{\rm bol}-0.68\log\mbox{Mass} 
              -3.48\log T_{e}+11.502.  
\label{eq:01}
\end{equation}
It can be made into an equation, $P(L,T_{e})$, for the lines of
constant period in the HR diagram once a mass-luminosity relation for
Cepheids is used to eliminate mass from equation~(\ref{eq:01}).  

     Observational determinations of many Cepheid masses are not
available, and we must rely on theoretical mass values from calculated
evolution tracks that pass through the instability strip. A summary of
such tracks is given in Tables 1 to 5 of \citeauthor*{SBT:99} for
tracks calculated from the Geneva models, in Table~11 for the Padua
tracks, and Table~12 for the Saio-Gautschy tracks.  
Detailed references for these models are in \citeauthor*{SBT:99}. 
The models of \citet{Marconi:etal:05} for solar metallicity and by
\citet{Bono:etal:00} for lower metallicities were also studied. 

     From all the models, normalized at $\log \mbox{Mass}=0.84$ at 
$\log L=3.80$, we have adopted
\begin{equation}
   \log\mbox{Mass} = 0.300\log L_{\rm bol}  - 0.300 
\label{eq:02}
\end{equation}
from the tracks. This is everywhere within 
$\Delta\log\mbox{Mass}=0.03\dex$ of the Geneva and Padua tables in
\citeauthor*{SBT:99} for all metallicities. 

     Putting equation~(\ref{eq:02}) into equation~(\ref{eq:01}) gives
the equation of the lines of constant period to be 
\begin{equation}
     \log L_{\rm bol} = 5.472 \log T_{e}+1.572\log P -18.406.
\label{eq:03}
\end{equation}
This produces a family of lines in the $\log L$, $\log T_{e}$  HR
diagram as $\log P$ is varied.

\begin{figure}[t]
   \plotone{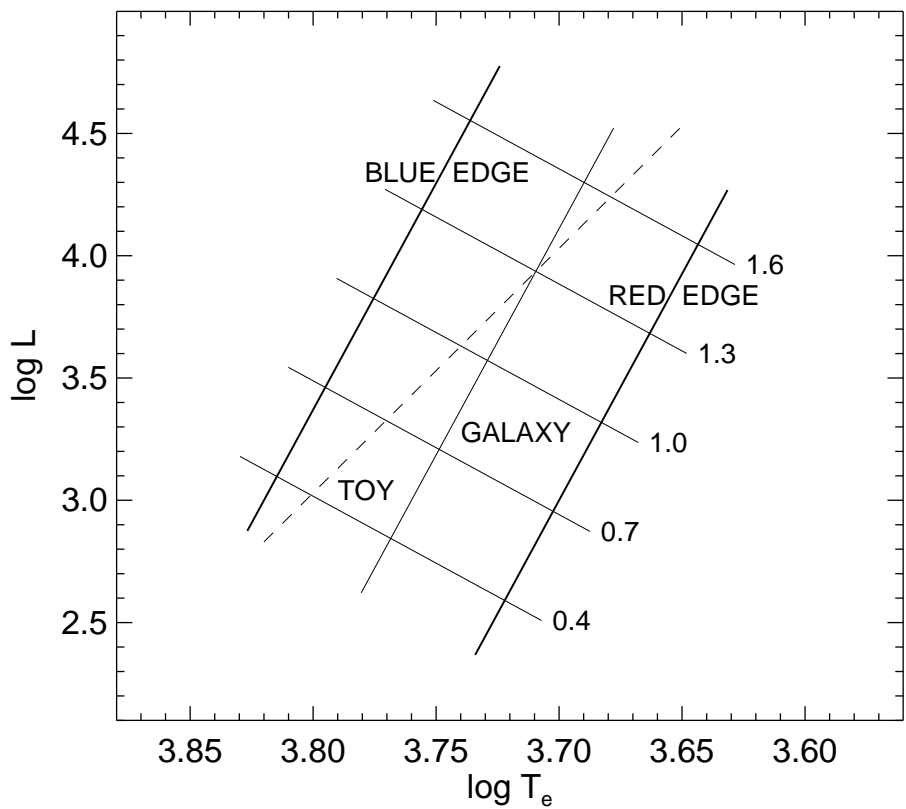}
   \caption{Schematic HR diagram in the vicinity of the Cepheid
     instability strip. The central line is the observed ridge line
     for the Galaxy taken from Figure~20 of STR\,04
     whose equation is $\log T_{e}=-0.054\log L_{V} + 3.922$. The
     dashed line is for a toy galaxy whose ridge-line equation is 
     $\log T_{e}=-0.100\log L_{V} + 4.103$. The borders of the
     instability strip are put parallel to the Galaxy ridge
     line. Lines of constant period, calculated from 
     equation~(\ref{eq:03}), are marked with their $\log P$ values 
     (in days).} 
\label{fig:01}
\end{figure}

     Figure~\ref{fig:01} shows such a family for $\log~P$ values of 
0.4, 0.7, 1.0, 1.3, and 1.6. The ridge-line instability strip for the
Galaxy is shown, using its equation of 
$\log T_{e}= -0.054\log L+3.922$ from \citeauthor*{STR:04}, Figure~20. 
The blue and red strip borders are arbitrarily drawn parallel to the
Galaxy ridge line using a temperature width of $\Delta\log T_{e}=0.06$ 
from the Galaxy ridge line. This is slightly wider than is observed 
(Fig.~20 of \citeauthor*{STR:04}), but is drawn to accommodate the
dashed strip line of a toy galaxy shown with the equation 
$\log T_{e}=- 0.100\log L+4.103$, similar to the instability strip of
LMC (again Fig.~20 of \citeauthor*{STR:04}), but drawn here without
the break at 10 days. The toy galaxy strip (the dashed line) has been
made to intersect the Galaxy strip at $\log P=1.3$ to insure that the
separate P-L relations also cross at this period.    


     The ridge-line P-L relations are obtained in an obvious way by
reading the $\log L$ (ordinate) values at the intersections of the
instability strip with the constant period lines for both the Galaxy
and the toy model. The fact that the resulting ridge-line P-L relation
obtained for the Galaxy in this way differs from that of the toy model
because of the different slopes of the instability strip ridge lines
is obvious from this construction. 

     For $\log L<4.0$, the instability strip has higher temperatures
for the toy model than for the Galaxy at a given period. Hence, the
intersection of the ridge-line strip with the constant period lines
occurs at brigher lumimosities for the toy than for the Galaxy, giving
a P-L relation for the toy model that is brighter than for the Galaxy
for all periods smaller than  $\log P=1.3$. The opposite is true for 
$\log P>1.3$. Hence the P-L relations will have different slopes, as
was to be shown.  

\begin{figure}[t]
   \plotone{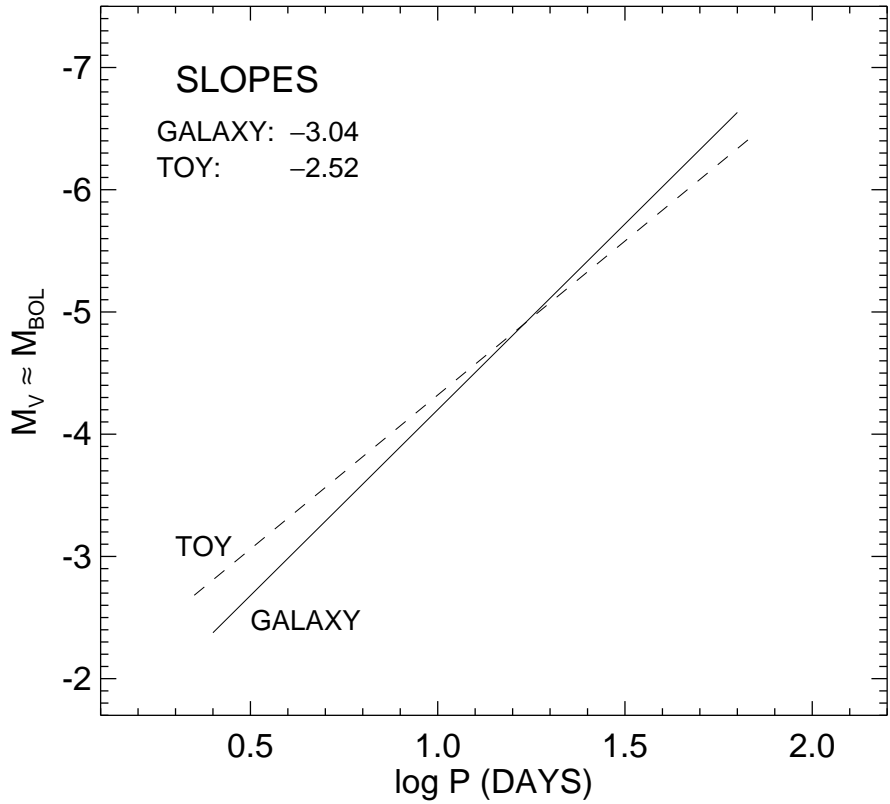}
   \caption{The two P-L relations for the two ridge lines in 
     Figure~\ref{fig:01}, determined from the intersections of the
     ridge lines of the Galaxy and the toy galaxy with the lines of
     constant period in Figure~\ref{fig:01}. The absolute magnitudes
     along the ordinate are transferred from Figure~\ref{fig:01} by 
     $M_{V}=-2.5\log L_{\rm bol}+4.75$ where the bolometric correction
     in $V$ is adopted to be zero.}  
\label{fig:02}
\end{figure}

     The discussion here in words could complete the promised
demonstration that the slope of the P-L relation is rigidly coupled
with the slope of the instability strip ridge line in the HR
diagram. However, to make the point more explicit, even to the point
that the discussion becomes unnecessarily more elementary, bordering
on pedantry, Figure~\ref{fig:02} displays the two different P-L
relations obtained by reading Figure~\ref{fig:01} in this way. 
The slope values for the Galaxy and the toy are marked in the Figure,
based on the adopted instability equations adopted for 
Figure~\ref{fig:01}. These slopes are similar to the actual slope
values measured for the Galaxy and the SMC from 
\citeauthor*{STR:04} (their eq.~[17]) and 
\citet{TSR:08}, hereafter \citeauthor*{TSR:08} (their eq.~[5]), and
set out again in Table~\ref{tab:01} here later.


\section{THE ALGEBRAIC SOLUTION USING DATA FROM THE GALAXY AND THE LMC}
\label{sec:03}
We can apply the pulsation equation directly to show the algebraic
solution for the same problem using real data, both for the equations
of the instability strips of the Galaxy and LMC and  the observed P-L
relations. The demonstration made here uses the equations for observed
P-L relations from \citeauthor*{STR:04} in their equation~(17) for the
Galaxy and their equations~(12) and (13) for the LMC. These are put
into equation~(\ref{eq:01}), which, together with the adopted
mass-luminosity equation~(\ref{eq:02}), gives a predicted 
$\log T_{e}$, $\log L$ instability ridge-line relation. This predicted
line is then compared with the observed instability strip equations
shown in Figure~20 of \citeauthor*{STR:04}.  

     We have used an explicit bolometric correction to change the 
$\log L_{V}$ values obtained from the observations into $\log L_{\rm bol}$
required in equations~(\ref{eq:01}) and (\ref{eq:02}), and back to 
$\log L_{V}$ to compare the predictions from the pulsation equation
with the observations. The bolometric corrections are interpolated
from Table~6 of \citeauthor*{SBT:99} for the appropriate metallicities
and surface gravities of the Cepheids. The turbulent velocity was
assumed to be $1.7\kms$. The surface gravities vary with radius, mass,
and  luminosity and therefore with period as a surrogate as 
$\log g = -1.09 \log P +2.64$ (eq.~[49] of \citeauthor*{STR:04}). The
metallicities are assumed to be [A/H]$=0.00$ for the Galaxy and $-0.5$
for LMC. The mass is from equation~(\ref{eq:02}). The obvious
arithmetic is not shown.      

     The resulting predictions of the instability strip ridge-line
equations are these:                 
\begin{equation}
  \log T_{e}(\mbox{predicted})  =  -0.040 \log L_{V}   + 3.854 
\label{eq:04}
\end{equation}
for the Galaxy at all periods, and, 
\begin{equation}
  \log T_{e}(\mbox{predicted})  =  -0.056 \log L_{V}   + 3.941 
\label{eq:05}
\end{equation}
for $P<10\;$days for the LMC, and   
\begin{equation}
  \log T_{e}(\mbox{predicted})  =  -0.081 \log L_{V}   + 4.020 
\label{eq:06}
\end{equation}
for $P>10\;$days, also for the LMC. 

     Note that the break in the $T_{e}-L$ instability strip relation
at $P=10\;$days in equations~(\ref{eq:05}) and (\ref{eq:06}) is
mirrored in the break in the P-L LMC relations given in equations~(12)
and (13) of \citeauthor*{STR:04}, and shown as Figure~4 there.

     For comparison with the predictions in 
equations~(\ref{eq:04})--(\ref{eq:06}) here, the {\em observed\/} ridge
lines of the strips in the Galaxy and the LMC, taken from the insert
equations shown in Figure~20 of \citeauthor*{STR:04}, are   
\begin{equation}
  \log T_{e}(\mbox{observed})  =  -0.054 \log L_{V}   + 3.922 
\label{eq:07}
\end{equation}
for the Galaxy at all periods, and
\begin{equation}
  \log T_{e}(\mbox{observed})  =  -0.050 \log L_{V}   + 3.936 
\label{eq:08}
\end{equation}
for $P<10\;$days, and 
\begin{equation}
  \log T_{e}(\mbox{observed})  =  -0.078 \log L_{V}   + 4.029 
\label{eq:09}
\end{equation}
for $P>10\;$days for the LMC.

     The near agreement of the predicted slopes of the instability
strips in equations (\ref{eq:04})--(\ref{eq:06}) with the observed
slopes in equations (\ref{eq:07})--(\ref{eq:09}) is the demonstration
we are seeking.   

     The agreement is good, but there is a disagreement in the 
temperature zero points between equations (\ref{eq:04})--(\ref{eq:06})
and equations (\ref{eq:07})--(\ref{eq:09}) by 
$\Delta \log T_{e}=0.018\dex$. The predicted temperatures are cooler 
than those observed. However, the difference is remarkably small, 
given the approximations we have made in the bolometric corrections, 
in the adopted temperature scale of \citeauthor*{SBT:99}, their Table~6,
and in the adopted van Albada-Baker theoretical zero point in 
equation~(\ref{eq:01}). 

     The temperature offset could be made zero if the zero point 
of the mass in equation~(\ref{eq:02}) would be made smaller by 
$0.09\dex$, but then the evolution mass would differ from the
pulsation mass by this amount. This is the expression of the previous
well known  mass ``problem'' which is solved here by the temperature
shift.  

     In this regard, it is useful to remark that many of the 
temperature scales in the current literature, for example as 
summarized by \citet{Sekiguchi:Fukugita:00} or by 
\citet{Cacciari:etal:05}, and including the one in 
\citeauthor*{SBT:99} that we have used here, differ among themselves
by as much as $0.025\dex$ in $\log T_{e}$ at fixed $B\!-\!V$. This,
then, is the temperature uncertainty in the temperature zero point in
Figure~20 of \citeauthor*{STR:04}. Our shifting of the predicted
temperature relative to the observed temperatures in
Figure~\ref{fig:03} by $0.018\dex$ is not excessive.  


     The observed (solid lines) and the predicted (dashed lines
shifted by $0.018\dex$ in $\log T_{e}$) instability strips for the
Galaxy and the LMC are shown in Figure~\ref{fig:03}. The agreement is
satisfactory, showing again that differences in the instability strip
loci causes differences in the slopes of the P-L relations. Hence 
the claims in the current literature, cited in the Introduction,
that a universal slope exists for the Cepheid P-L relation are
inconsistent with Figures~\ref{fig:01}--\ref{fig:03} which show
different positions of the instability strip in different galaxies.

\section{SUMMARY OF OBSERVED P-L SLOPE DIFFERENCES IN SELECTED GALAXIES}
\label{sec:04}
The arguments given in the previous sections rely on knowledge
of the temperatures of the instability strips. These can only be
measured using reddening corrected colors, and these are reliable only
if the reddening of the individual Cepheids can be determined by some
method other than by using a fiducial  period-color (P-C)
relation. The reason is that if the temperatures of the instability
strips differ from galaxy-to-galaxy, presumably because of chemical
composition differences, the P-C relations will also differ. There
will be no correct fiducial P-C template from which to determine the
reddening if the chemical compositions vary greatly. The intrinsic P-C
relations will differ from galaxy-to-galaxy depending on the chemical
composition, and the reddenings are therefore indeterminate.

     Presently, it is only the Galaxy, LMC, and SMC that can be
subjected to the analysis given here because it is only for these
galaxies that the reddening of their Cepheids have been determined by
methods other than by comparing with some adopted fiducial P-C
relation.

     However, for some galaxies with enough Cepheids, and where the
differential reddening between the Cepheids is small enough to be
ignored, comparison of the P-L slopes can be made directly from the
data. The result for the Galaxy, NGC\,3351, NGC\,4321, LMC, SMC,
IC\,1613, NGC\,3109, and Sextans A and B is shown in
Figure~\ref{fig:04}. The adopted data for the P-L relations are in
Table~\ref{tab:01}. The equations for the apparent magnitude and
absolute magnitude P-L relations are $V^{0}= a\log P + b$, and 
$M_{V}^{0} = a\log P + c$. Column~(2) is the $\log$ of the
oxygen-to-hydrogen ratio from Table~4 of \citeauthor*{TSR:08}.   
Column~(3) shows the observed value of $a$, which is the slope of the 
apparent magnitude P-L relation taken from the same sources (but
changed slightly in a few entries here) as were used for Table~4 of
\citeauthor*{TSR:08}. Column~(4) lists the apparent magnitude P-L
intercept, $b$, as observed. Column~(5) lists the $(m-M)^{0}$
distance modulus adopted in \citeauthor*{TSR:08}. The absolute
magnitude P-L relation is in column~(6), which is column~(4) minus
column~(5). The literature source is in column~(7). The 
resulting P-L relations, calculated from the $a$ and $c$ values in
Table~\ref{tab:01}, are shown in Figure~\ref{fig:04}. The slopes for
the NGC\,3351/ NGC\,4321 combination and the Galaxy are the steepest
of those shown, and are similar. That of the LMC is next steepest.  


     The slope of the Galaxy P-L relation in \citeauthor*{TSR:03} and
\citeauthor*{STR:04} is based on averaging the results using the
moving atmosphere method (the Baade-Becker-Wesselink procedure) and
the independent main  sequence fitting method. Nevertheless, the
resulting slope of the P-L slope has been questioned as being too steep 
\citep[cf.][]{Gieren:etal:05b,vanLeeuwen:etal:07}. 
However, the slopes of the NGC\,3351 and NGC\,4321 combined P-L 
relation, and that of M31 by \citet{Vilardell:etal:07} 
are equally steep as for the Galaxy. The M31 slope by
\citeauthor{Vilardell:etal:07} has been redetermined by 
\citeauthor*{TSR:08}.  
The original slope by Villardel and collaborators was based on 
$E(B\!-\!V)$ values using the LMC P-C relation rather than the more 
correct higher metallicity P-C relation for the Galaxy. The       
resulting $E(B\!-\!V)$ values turns out to depend on period as a 
further complication. But even discounting the M31 case, the 
steep slope for NGC\,3351 and NGC\,4321 from \citeauthor*{TSR:08} 
(their Fig.~2), supports the Galaxy slope that we derived in
\citeauthor*{STR:04} and its difference from the P-L slope in LMC.        

     The strongest evidence for the difference as function of
metallicity is the data for NGC\,3109 \citep{Pietrzynski:etal:06} 
which has the well determined P-L slope of $dM_{V} /d\log P = -2.13$. 
This differs significantly from the slopes of either the Galaxy 
or NGC\,3351/ NGC\,4321 at $-3.10$, or $-2.92$ for M31, and even for
the LMC at $-2.70$. The six longest period Cepheids in NGC\,3109 with 
$\log P > 1.3$ are too faint by  $\sim\!0.2\mag$ 
(Fig.~4 of \citealt{Pietrzynski:etal:06}) compared with either the
Galaxy or the LMC P-L relations.   

\begin{figure}[t]
   \plotone{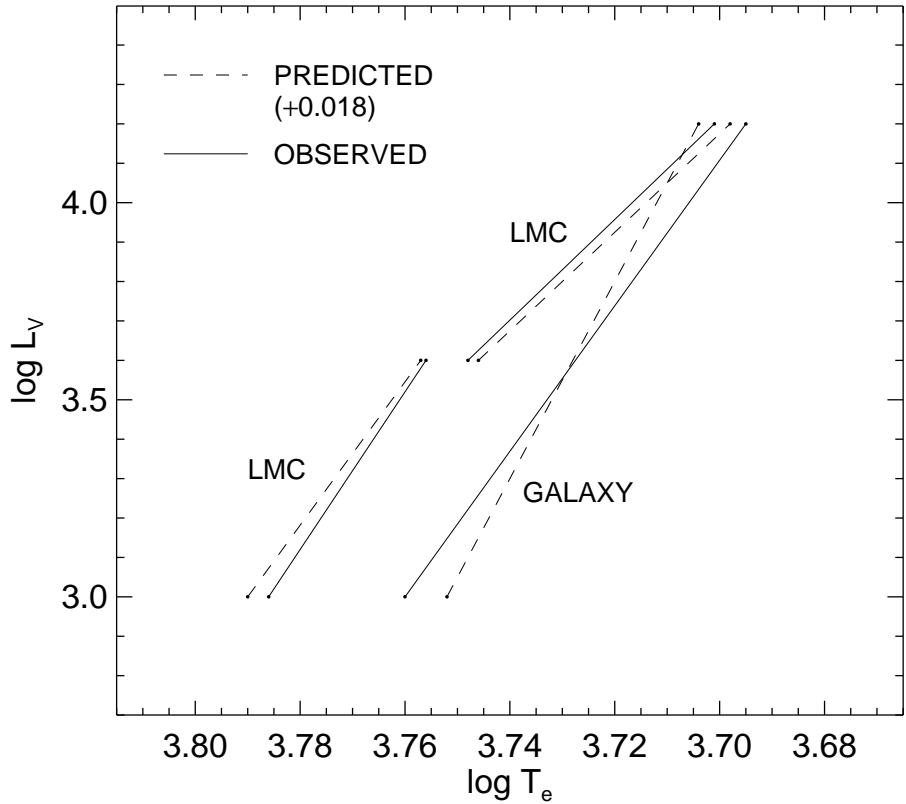}
   \caption{The algebraic demonstration of the rigid coupling between
     the slopes of the instability strip and the slope of the P-L
     relation required by the pulsation equation. Predicted (dashed
     lines) slopes and zero points for these instability strip ridge
     lines in the Galaxy and the LMC are compared with the observed
     (solid) lines from Figure~20 of \citeauthor*{STR:04}. The
     predictions are made by inserting the equations of the observed
     P-L relations for the Galaxy and the LMC into the pulsation
     equation~(\ref{eq:01}). The predicted zero points are moved by
     0.018 in $\log T_{e}$, hotter.}   
\label{fig:03}
\end{figure}

\begin{figure}[t]
   \plotone{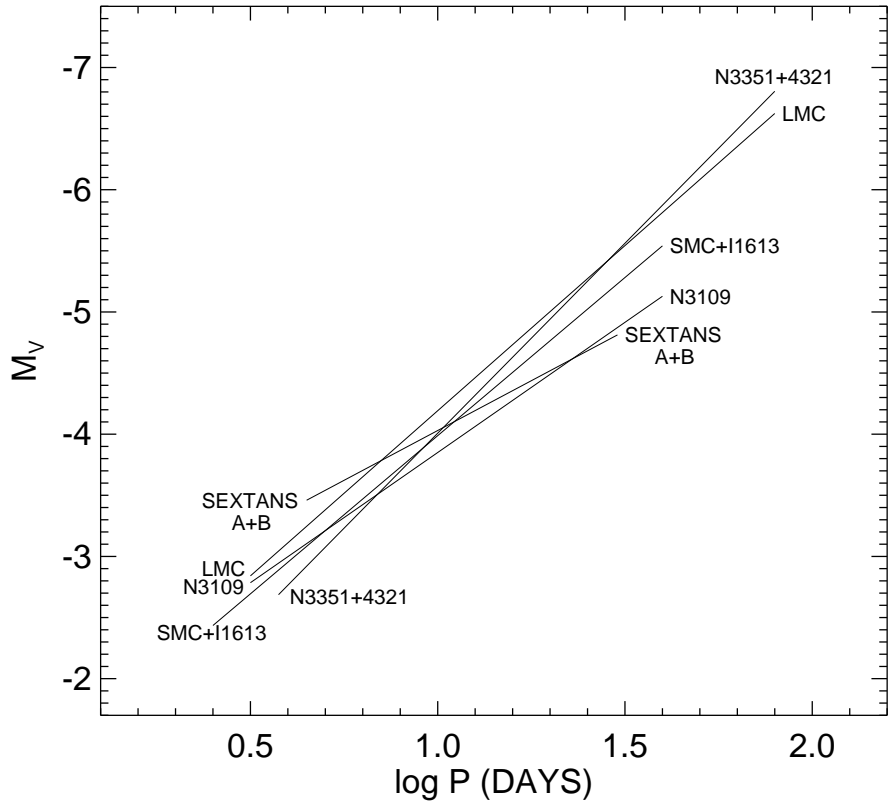}
   \caption{The observed ridge lines of the P-L relations for eight
     galaxies listed in Table~\ref{tab:01}. The P-L relation for the
     Galaxy (not shown) is nearly identical with the combined
     NGC\,3351 and NGC\,4321 line, and has the steepest slope. 
     The agreement between the Galaxy and the combined NGC\,3351 and
     NGC\,4321 slopes argues for the correctness of the steep slope
     for the Galaxy P-L relation.}
\label{fig:04}
\end{figure}

     Figure~\ref{fig:04}, similar in principle to Figure~5 in
\citeauthor*{TSR:08}, together with Figure~\ref{fig:03} here, is our
chief case for non-unique P-L relations between galaxies of different
chemical compositions. The complications that this portends for
determining the scale of extragalactic distances from Cepheids to
within $\sim\!15\%$, unless special corrections for the difference
are applied, is discussed elsewhere  
(cf. \citealt{STT:06}; \citealt{STS:06}; \citeauthor*{TSR:08}).

\acknowledgments
It is a pleasure to thank Bernd Reindl for his skill in the 
preparation of the diagrams and the text for publication. We 
also thank John Grula, Carnegie editorial chief, for his 
liaison with the press.



\clearpage


\begin{deluxetable}{lclccccc}
\tablewidth{0pt}
\tabletypesize{\footnotesize}
\tablecaption{{\sc Observed (P-L)$_{V}$ Relations for Ten Galaxies with
  Different Chemical Compositions}\label{tab:01}}   
\tablehead{
 \colhead{Name}        & 
 \colhead{[O/H]}       &    
 \colhead{$a$}         &  
 \colhead{$b$}         & 
 \colhead{$(m-M)^{0}$} &
 \colhead{$c$}         &    
 \colhead{Ref}               
\\
 \colhead{(1)}         & 
 \colhead{(2)}         & 
 \colhead{(3)}         & 
 \colhead{(4)}         & 
 \colhead{(5)}         & 
 \colhead{(6)}         & 
 \colhead{(7)}     
} 
\startdata
Galaxy         & $8.60$ & $-3.087$ & \nodata & \nodata & $-0.91$ & 1 \\
NGC\,3351/4321 & $\langle8.80\rangle$ & $-3.108$ & mean & mean & $-0.90$ & 2 \\
M31            & $8.66$ & $-2.92$  & \nodata & $24.43$ & \nodata & 3 \\
LMC            & $8.34$ & $-2.702$ & $17.05$ & $18.54$ & $-1.49$ & 4 \\
SMC            & $7.98$ & $-2.588$ & $17.53$ & $18.93$ & $-1.40$ & 5 \\
IC\,1613       & $7.86$ & $-2.698$ & $23.08$ & $24.35$ & $-1.27$ & 6 \\
NGC\,3109      & $8.06$ & $-2.130$ & $23.73$ & $25.45$ & $-1.72$ & 7 \\
Sextans A/B    & $7.52$ & $-1.628$ & $23.10$ & $25.80$ & $-2.40$ & 8 \\
\enddata
\tablerefs{
 (1) \citeauthor*{STR:04}, eq.~[17]; 
 (2) \citeauthor*{TSR:08}, Fig.~2; 
 (3) \citealt{Vilardell:etal:07}; 
 (4) \citeauthor*{STR:04}, eq.~[8]; 
 (5) \citeauthor*{TSR:08}, eq.~[5]; 
 (6) \citealt{Antonello:etal:06}; 
 (7) \citealt{Pietrzynski:etal:06}, Fig.~4; 
 (8) \citealt{Piotto:etal:94}.
} 
\end{deluxetable}

\clearpage


\epsscale{0.8} 

\clearpage

\clearpage

\clearpage

\end{document}